\documentclass[aps,prl,reprint,superscriptaddress]{revtex4-1}

\usepackage[english]{babel}
\usepackage{graphicx}
\usepackage{amssymb}    	
\usepackage{gensymb}		
\usepackage{siunitx}
\usepackage{color,soul}
\DeclareGraphicsExtensions{.pdf,.png}

\begin{document}


\title{Modified Archimedes' principle predicts rising and sinking of intruders in sheared granular flows}

\author{Lu Jing}
\affiliation{Department of Chemical and Biological Engineering, Northwestern University, Evanston, IL 60208, USA}

\author{Julio M. Ottino}
\affiliation{Department of Chemical and Biological Engineering, Northwestern University, Evanston, IL 60208, USA}
\affiliation{Department of Mechanical Engineering, Northwestern University, Evanston, IL 60208, USA}
\affiliation{Northwestern Institute on Complex Systems (NICO), Northwestern University, Evanston, IL 60208, USA}

\author{Richard M. Lueptow}
\email[]{r-lueptow@northwestern.edu}
\affiliation{Department of Chemical and Biological Engineering, Northwestern University, Evanston, IL 60208, USA}
\affiliation{Department of Mechanical Engineering, Northwestern University, Evanston, IL 60208, USA}
\affiliation{Northwestern Institute on Complex Systems (NICO), Northwestern University, Evanston, IL 60208, USA}

\author{Paul B. Umbanhowar}
\affiliation{Department of Mechanical Engineering, Northwestern University, Evanston, IL 60208, USA}

\date{\today}

\begin{abstract}
We computationally determine the force on single spherical intruder particles in sheared granular flows as a function of particle size, particle density, shear rate, overburden pressure, and gravitational acceleration. The force scales similarly to, but deviates from, the buoyancy force predicted by Archimedes' principle. The deviation depends only on the intruder to bed particle size ratio, but not the density ratio or flow conditions. We propose a simple force model that successfully predicts whether intruders rise or sink, knowing only the size and density ratios, for a variety of flow configurations in physical experiments. 
\end{abstract}


\maketitle

Intruder particles in fluidized or flowing granular beds tend to segregate (rise or sink) due to their size or density difference with the bed particles \cite{duran_arching_1993,knight_vibration-induced_1993,shinbrot_reverse_1998,mobius_brazil-nut_2001,shishodia_particle_2001,hong_reverse_2001,breu_reversing_2003,huerta_vibration-induced_2004,huerta_archimedes_2005,tripathi_numerical_2011,van_der_vaart_underlying_2015,guillard_scaling_2016,jing_micromechanical_2017,van_der_vaart_segregation_2018,staron_rising_2018}. 
Segregation in vibrofluidized systems, the Brazil nut effect \cite{mobius_brazil-nut_2001}, depends on various mechanisms \cite{duran_arching_1993,knight_vibration-induced_1993,shinbrot_reverse_1998,mobius_brazil-nut_2001,shishodia_particle_2001,hong_reverse_2001,breu_reversing_2003,huerta_vibration-induced_2004,huerta_archimedes_2005} including buoyancy. With sufficient fluidization, the buoyancy force on an intruder follows Archimedes' principle \cite{shishodia_particle_2001,huerta_archimedes_2005}, thus explaining the phase transition between normal and reverse Brazil nut effects \cite{hong_reverse_2001,breu_reversing_2003}. In contrast to this clear picture, the force driving segregation in \emph{sheared} granular flows remains elusive. While extensive research has focused on segregation of flowing bidisperse mixtures from the continuum perspective \cite{gray_particle_2017,umbanhowar_modeling_2019}, quantitative studies of the particle-scale segregation force are fewer and more recent. \citet{guillard_scaling_2016} proposed a virtual spring based force meter in numerical simulations that allows direct measurement of the segregation force in shear flows. They interpreted the force as summed contributions from normal and shear stress gradients. Van der Vaart \emph{et al.} \cite{van_der_vaart_segregation_2018} applied a similar approach in chute flows and decomposed the measured force into lift and buoyancy-like forces. Despite these insights, a generalized characterization of the segregation force is still lacking in either size \cite{van_der_vaart_underlying_2015,guillard_scaling_2016,jing_micromechanical_2017,van_der_vaart_segregation_2018,staron_rising_2018} or density \cite{tripathi_numerical_2011} segregation, as well as more complicated situations of combined size and density segregation. For example, \citet{felix_evidence_2004} found an interplay between size and density whereby segregation can change direction (rise or sink) more than once with the monotonic increase of intruder size. This raises the question whether intruder segregation is predictable knowing only the size and density differences.

This Letter solves the puzzle by providing a generalized force model that allows shear-induced segregation to be viewed as a result of the imbalance between the gravitational force and a \emph{modified} Archimedes buoyancy force. The model successfully predicts segregation transitions in various experiments over a wide range of size and density ratios, which enhances prediction of size and density segregation in industrial and geophysical granular flows.

\begin{figure}[b!]
\includegraphics{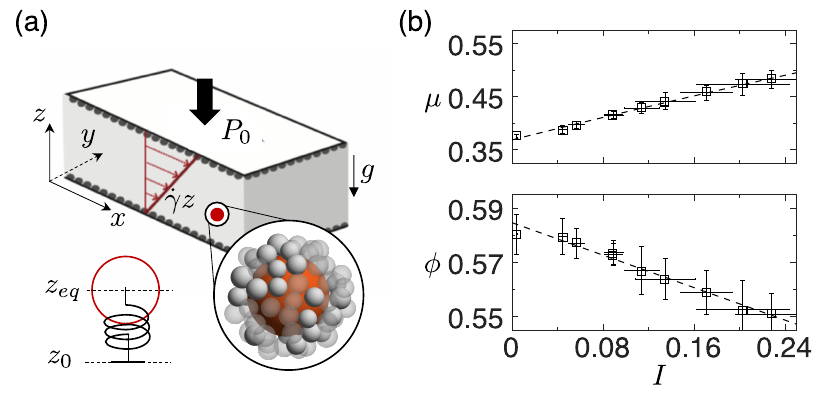}%
\caption{\label{fig:1}
	(a) Intruder particle (red) in a sheared granular bed. A virtual spring measures the vertical force on the intruder.
	(b) Rheology: $\mu$ and $\phi$ vs.\ $I$. Error bars indicate $\pm1$ standard deviation. Dashed lines are fits to the $\mu(I)$ rheology \cite{jop_constitutive_2006}.
	}
\end{figure}

\emph{Methods.---}We simulate single spherical intruders in sheared granular flows using a discrete element method code \textsc{liggghts} \cite{kloss_models_2012}. As sketched in Fig.~\ref{fig:1}(a), bed particles of diameter $d$ and density $\rho$ are sheared in a streamwise ($x$) and spanwise ($y$) periodic box of length $30d$, height $30d$, and width $10d$ to $30d$ (varied as needed) in the presence of gravity ($g=9.81$~\si{m/s^2}). We use $d=5$~\si{mm}, with $10\%$ uniform size polydispersity to avoid layering, $\rho=2500$~\si{kg/m^3}, and the Hertz contact model with Young's modulus $5\times10^7$~\si{Pa}, Poisson's ratio $0.4$, restitution coefficient $0.8$, and friction coefficient $0.5$. The top and bottom walls are roughened with randomly distributed stationary particles to prevent slippage \cite{jing_characterization_2016}, and an overburden pressure $P_0$ is applied on the top wall. The shear flow is driven by moving the top wall and applying a stabilizing force to each particle in the $x$-direction \cite{fry_effect_2018}; at each time step, for a particle with streamwise velocity $u_p$ and vertical position $z_p$, a small force proportional to $\dot{\gamma}z_p-u_p$ is added to maintain a linear velocity profile $\dot{\gamma}z$ across all particles [Fig.~\ref{fig:1}(a)]. This allows us to conveniently generate a wide range of shear flows with $600~\si{Pa}\leqslant P_0\leqslant3000~\si{Pa}$ and $1~\si{s^{-1}}\leqslant\dot{\gamma}\leqslant40~\si{s^{-1}}$. As Fig.~\ref{fig:1}(b) shows, the inertial number $I=\dot{\gamma}d\sqrt{\rho/\sigma_{zz}}$ ranges from $0.005$ to $0.25$, where $\sigma_{zz}$ is the vertical normal stress, and the effective friction $\mu$ and packing fraction $\phi$ follow the $\mu(I)$ rheology \cite{jop_constitutive_2006}. 

An intruder of diameter $d_i$ and density $\rho_i$ is placed near the middle of the bed (initial height $z_0$), with size ratio $R=d_i/d$ varying from $0.5$ to $8$ and density ratio $R_\rho=\rho_i/\rho$ varying from $0.5$ to $3$. The same streamwise stabilizing force applies to the intruder. To measure the vertical force driving segregation, we follow \citet{guillard_scaling_2016} and tether the intruder to a vertical spring (leaving free the other five degrees of freedom), which causes it to fluctuate about an equilibrium height $z_{eq}$ [Fig.~\ref{fig:1}(a)]. In steady state, the net contact force exerted on the intruder by the neighboring bed particles, the bed force $F$, is balanced by the spring force and the gravitational force, i.e., $F=k_{sp}(z_{eq}-z_0)+m_ig$, where $k_{sp}$ is the virtual spring stiffness and $m_i$ is the intruder mass. The spring acts as a virtual force meter and the measurement of $F$ is insensitive to $k_{sp}$ \cite{guillard_scaling_2016,van_der_vaart_segregation_2018}. Uncertainties (error bars) of $F$ are estimated considering temporal correlations \cite{zhang_calculation_2006} of the fluctuations of the intruder height about $z_{eq}$.

\begin{figure}[t!]
\includegraphics{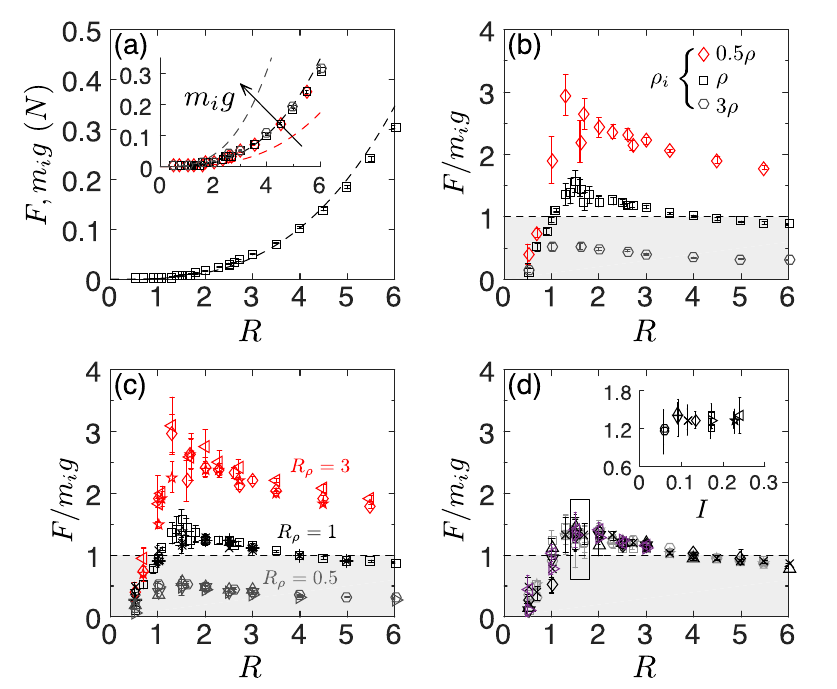}%
\caption{\label{fig:2} Dependence of $F$ on various parameters.
	(a) $F$ and $m_ig$ vs.\ $R$ for $\rho_i=\rho$ ($\rho=2500$~\si{kg/m^3}, $P_0=600$~\si{Pa}, $\dot{\gamma}=10$~\si{s^{-1}}, $g=9.81$~\si{m/s^2}). Inset: Influence of $\rho_i$ for $0.5\rho\leqslant\rho_i\leqslant3\rho$.
	(b) $F/m_ig$ vs.\ $R$ for $0.5\rho\leqslant\rho_i\leqslant3\rho$.
	(c) $F/m_ig$ vs.\ $R$ for $0.5\leqslant R_\rho\leqslant3$ with varying $\rho$ and $\rho_i$.
	(d) $F/m_ig$ vs.\ $R$ for $600~\si{Pa}\leqslant P_0\leqslant3000~\si{Pa}$, $10~\si{s^{-1}}\leqslant\dot{\gamma}\leqslant40~\si{s^{-1}}$, and $5~\si{m/s^2}\leqslant g \leqslant15~\si{m/s^2}$. Inset: $F/m_ig$ vs.\ $I$ for $1.4\leqslant R\leqslant1.6$ (selected for illustration). Shaded (unshaded) areas in (b--d) indicate that the intruder sinks (rises) from the initial position.
	}
\end{figure}

\emph{Results.}---Figure~\ref{fig:2}(a) shows that, for $\rho_i=\rho$, $F$ (symbols) and $m_ig$ (dashed curve) increase similarly with $R$. However, subtle differences between $F$ and $m_ig$ indicate imbalanced forces that drive segregation. To better visualize the differences, the ratio $F/m_ig$ is plotted in Fig.~\ref{fig:2}(b). Focusing on $\rho_i=\rho$, $F/m_ig$ is less than one for $R<1$, i.e., a small intruder is pulled down by gravity. As $R$ is increased above one, $F/m_ig$ becomes greater than one, i.e, a large intruder is pushed up by the bed force. These scenarios are consistent with typical percolation and squeeze expulsion explanations for size segregation \cite{savage_particle_1988,jing_micromechanical_2017}. Notably, $F/m_ig$ falls slightly below one for $R>4$, since $m_ig$ increases more rapidly than $F$ as $R$ increases; thus, very large intruders sink. Such reverse segregation has been reported \cite{felix_evidence_2004} but not yet quantitatively addressed \cite{guillard_scaling_2016,van_der_vaart_segregation_2018}.

Next, we vary $R_\rho$ by changing $\rho_i$. The inset of Fig.~\ref{fig:2}(a) shows that $F$ remains unchanged as $\rho_i$ increases from $0.5\rho$ to $3\rho$ (different symbols), whereas $m_ig$ obviously depends on $\rho_i$ (dashed curves). Therefore, the intruder density (and the weight) does not directly affect the bed force but alters segregation behavior by changing the ratio $F/m_ig$. As shown in Fig.~\ref{fig:2}(b), a sufficiently heavy intruder ($\rho_i=3\rho$) sinks regardless of its diameter, as the bed force can never support its weight; a light intruder ($\rho_i=0.5\rho$) rises for $R>1$, as its weight is less than the force pushing it upward.  

We also vary $R_\rho$ by changing $\rho$ at constant $\rho_i$ such that $m_ig$ remains the same but $F$ varies significantly. Combined with the data in Fig.~\ref{fig:2}(b), plots for different $\rho_i$ and $\rho$ collapse on curves distinguished only by $R_\rho$ [Fig.~\ref{fig:2}(c)], i.e., whether an intruder rises or sinks is determined only by the \emph{relative} diameter and density.

Finally, Fig.~\ref{fig:2}(d) shows that flow conditions $P_0$, $\dot{\gamma}$, and $g$ have no significant impact on $F/m_ig$ over a wide range of variation. As illustrated in Fig.~\ref{fig:2}(d) inset, $F/m_ig$ is essentially independent of $I$ for $0.05<I<0.25$, a range encompassing typical inertial flows \cite{azema_internal_2014}. Reducing $I$ toward the quasistatic limit (typically $10^{-3}$) may enhance the segregation force \cite{guillard_scaling_2016}, a point we address below.

\begin{figure}[t!]
\includegraphics{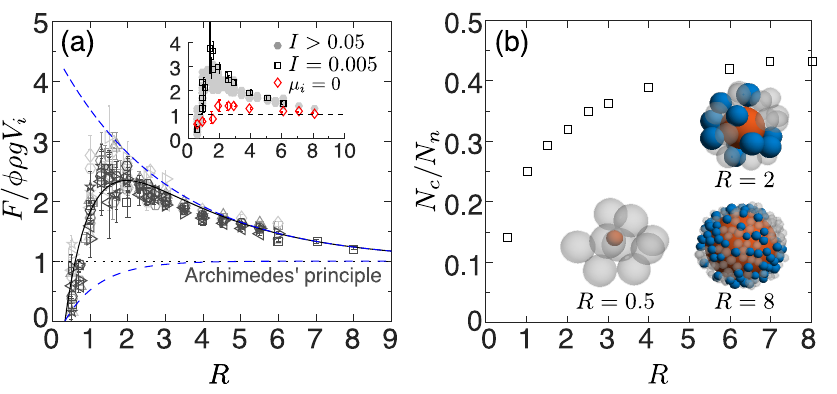}%
\caption{\label{fig:3} 
	(a) $F/\phi\rho gV_i$ vs.\ $R$, with varying $\rho_i$, $\rho$, $P_0$, $\dot{\gamma}$, and $g$. Solid curve is a fit to Eq.~(\ref{eq:F}). Dashed curves show the two exponential terms defining $f(R)$. Inset: extreme cases with $I=0.005$ ($P_0=3000$~\si{Pa}, $\dot{\gamma}=1$~\si{s^{-1}}) and $\mu_i=0$ ($P_0=1800$~\si{Pa}, $\dot{\gamma}=20$~\si{s^{-1}}), respectively.
	(b) $N_c/N_{n}$ vs.\ $R$ ($P_0=2200$~\si{Pa}, $\dot{\gamma}=20$~\si{s^{-1}}). Inset: $xz$-plane view of contact network at various $R$, showing intruders (red), contacting neighbor particles (blue), and noncontacting neighbor particles (gray).}
\end{figure}

\emph{Scaling.---}We now focus on the scaling of $F$ and test an Archimedes buoyancy-like force scale, $\phi\rho gV_i$, where $V_i$ is the intruder volume, viewing the flow as a ``fluid'' of bulk density $\phi\rho$ with normal stress gradient $\phi\rho g$. Figure~\ref{fig:3}(a) shows $F/\phi\rho gV_i$ vs.\ $R$ for $204$ distinct simulations. All data collapse on a master curve, confirming that $F$ scales with the buoyancy force. However, the master curve deviates from $F/\phi\rho gV_i=1$; it starts below one for $R<1$, increases and reaches a maximum of about $2.5$ at $R\approx2$, and approaches one as $R$ increases to large values.

The deviation between $F$ and $\phi\rho gV_i$ appears to originate in geometric effects at the particle level. For relatively large intruders [e.g., $R=8$ in Fig.~\ref{fig:3}(b) inset], a large number of contacting neighbor particles (blue) transmit contact stress in a nearly uniform manner, consistent with the fluid buoyancy analogy $F\approx\phi\rho gV_i$ \cite{van_der_vaart_segregation_2018}.
As $R$ is decreased to intermediate values (e.g., $R=2$), contact uniformity breaks down significantly; this is characterized by the time-averaged number ratio of contacting neighbor particles ($N_c$) to all ``nearby'' particles ($N_n$) defined within a distance $d+d_i/2$ from the intruder center, which decreases rapidly as $R$ decreases [Fig.~\ref{fig:3}(b)]. Consequently, noncontacting neighbor particles (gray) are more likely to lose connection in stress transmission, which in turn leads to more contact forces passing through the intruder and thus a higher net force compared to the uniform limiting case, i.e., $F>\phi\rho gV_i$. 
As $R$ is further decreased below one (e.g., $R=0.5$), brief collisions dominate \cite{silbert_rheology_2007} and the intruder tends to percolate through voids without enduring contacts \cite{jing_micromechanical_2017}, resulting in a net contact force smaller than the uniform limiting case, i.e., $F<\phi\rho gV_i$.

The geometric effects are associated with the frictional nature of granular contacts. For frictionless intruders [$\mu_i=0$ in Fig.~\ref{fig:3}(a) inset], $F/\phi\rho gV_i$ collapses toward one, explaining previous observations that large intruders do not rise with low friction \cite{jing_micromechanical_2017,van_der_vaart_segregation_2018}.
In nearly quasistatic flows [$I=0.005$ in Fig.~\ref{fig:3}(a) inset], $F/\phi\rho gV_i$ is higher likely due to enhanced frictional resistance to deformation near yielding \cite{kang_archimedes_2018}. This effect tends to plateau above yielding, explaining the insensitivity of $F/m_ig$ to $I$ in Fig.~\ref{fig:2}(d). A similar trend of enhanced segregation force only at very low $I$ was found in previous two-dimensional simulations \cite{guillard_scaling_2016}.


\emph{Model.---}The master scaling curve in Fig.~\ref{fig:3}(a) suggests a modified Archimedes' principle of the form

\begin{equation}\label{eq:F}
	F=f(R)\phi\rho gV_i,
\end{equation}

\noindent where $f(R)$ is a dimensionless scale factor.
Based on two geometric effects that dominate in different ranges of $R$, i.e., percolation-induced force weakening for $R<1$ and nonuniformity-induced force strengthening for $R>1$, we propose $f(R)=(1-c_1e^{-R/R_1})(1+c_2e^{-R/R_2})$, where $c_1$, $c_2$, $R_1$, and $R_2$ are fitting parameters. The first term [lower dashed curve in Fig.~\ref{fig:3}(a)] represents stronger percolation (thus smaller bed force) as $R$ decreases; its exponential form is chosen to reconcile the exponential dependency of percolation probability \cite{savage_particle_1988} and percolation velocity \cite{khola_correlations_2016} on $R$. The second term [upper dashed curve in Fig.~\ref{fig:3}(a)], which decreases toward one as $R$ increases, accounts for decreased uniformity of contacts around the intruder at small $R$ [Fig.~\ref{fig:3}(b)]. Fitting to the data in Fig.~\ref{fig:3}(a) gives $c_1=1.43$, $c_2=3.55$, $R_1=0.92$, and $R_2=2.94$, where $R_1$ and $R_2$ are characteristic size ratios for the two effects to dominate. The two terms together recover the continuum argument, $f(\infty)\rightarrow1$, and the force balance in monodisperse flows, $f(1)=1/\phi$ (i.e., $F=\rho gV_i$ at $R=1$). Although $\phi$ is case specific, the fitting results in $\phi=1/f(1)=0.55$, a value agreeing with Fig.~\ref{fig:1}(b), which further supports the model.



\begin{figure}[b!]
\includegraphics{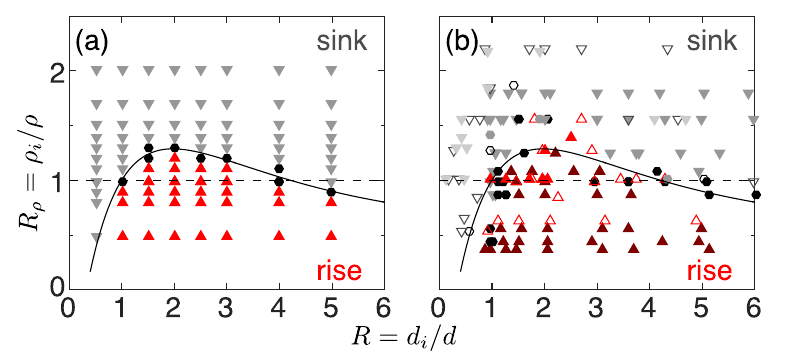}%
\caption{\label{fig:4} Segregation transition predicted by Eq.~(\ref{eq:F}) compared to (a) $88$ simulations ($P_0=600$~\si{Pa}, $\dot{\gamma}=20$~\si{s^{-1}}) and (b) $189$ experiments \cite{felix_evidence_2004}. Curves show $R_\rho=\phi f(R)$ with $\phi=0.55$. Up(down)-pointing triangles indicate rising (sinking) intruders; circles indicate neutral intruders. Data in (b) are from rotating drums (darker filled symbols), chute flows (lighter filled symbols), and heap flows (unfilled symbols).}
\end{figure}

Despite the empirical formulation of $f(R)$, the model adopts a minimum number of parameters to describe the data over the full range of $R$, clearly indicates two geometric effects, each associated with physically reasonable characteristic size ratios, and is appropriately constrained by limiting cases. Moreover, it provides a simple means to predict segregation based only on $R$ and $R_\rho$. An intruder in a sheared bed is ``neutral'' when the bed force $f(R)\phi\rho gV_i$ offsets its weight $\rho_i gV_i$, i.e., $R_\rho=\phi f(R)$, which describes a curve dividing the $R$-$R_\rho$ space into ``rise'' (below the curve) and ``sink'' (above the curve) zones; see Fig.~\ref{fig:4}. To validate this phase diagram, we simulate single \emph{untethered} intruders with varying $d_i$ and $\rho_i$, observing whether they rise, sink, or neither (i.e., mean displacement less than $3d$) over $500$~\si{s} of simulation. The predictions are in excellent agreement with the simulation results [Fig.~\ref{fig:4}(a)].

To further demonstrate the generality of the segregation transition predicted by Eq.~(\ref{eq:F}), we compare it with experiments by \citet{felix_evidence_2004}, who studied segregation of tracer particles of different sizes and densities in various configurations, i.e., rotating drums, chute flows, and heap flows. Despite the different flow geometries, the segregation direction in the experiments agrees remarkably well with the predictions of our phase diagram [Fig.~\ref{fig:4}(b)], showing the capability of our model to predict segregation for varying size and density ratios as well as different flow conditions. The few mismatches occurring near the neutral curve are mainly from chute and heap flows, where only loose criteria for the segregation direction were applied in the experiments \cite{felix_evidence_2004}.

\emph{Discussion}.---Our segregation force model respects the continuum limit ($R\gg1$) in that whether an intruder rises or sinks depends only on its density relative to the surrounding flow ($\rho_i/\phi\rho$), noting $f(\infty)\rightarrow1$. For intruders somewhat larger than the bed particles ($R>1$), discrete particle interactions result in a \emph{positive} deviation from Archimedes' principle, an extra lift effect underlying the rise of large particles in many size segregation studies \cite{gray_particle_2017,umbanhowar_modeling_2019}. The maximum deviation at $R\approx2$ explains the optimal segregation rate at $R\approx2$ and the saturation of segregation velocity for $R>2$ \cite{golick_mixing_2009,schlick_modeling_2015,jones_asymmetric_2018}. The modified Archimedes' principle described here bridges segregation mechanisms in noncontinuum situations with the continuum buoyancy force, suggesting a unifying framework for understanding forces in granular media. Indeed, Archimedes' principle with appropriate corrections applies to dense granular shear flows (this work), creeping granular fluids \cite{nichol_flow-induced_2010}, vibrofluidized granular gases \cite{huerta_vibration-induced_2004,huerta_archimedes_2005}, and plastic granular solids near yielding \cite{kang_archimedes_2018}. 



The model proposed here enables prediction of intruder segregation for various flow configurations based only on size and density ratios.
The finding that $F/\phi\rho gV_i$ is insensitive to external flow conditions is not to be confused with the known effects of shear rate and confining pressure on segregation \emph{velocity} \cite{fry_effect_2018,fan_modelling_2014,schlick_modeling_2015,liu_transport_2017}. While the direction of segregation is determined by competition between the bed force and the gravitational force, the segregation velocity depends further on resistive forces (often viewed as drag). Understanding the drag force has proved challenging due to the difficulty in isolating driving and drag terms from contact forces \cite{tripathi_numerical_2011,staron_rising_2018,duan_kinetic_2019}. Now with the generalized driving force model we provide, it is possible to calculate the drag force on moving intruders. It is also relevant to consider varying the particle species concentration around the tethered intruders to account for general industrial and geophysical settings \cite{gray_theory_2005,tunuguntla_mixture_2014,hill_segregation_2014,gray_particle-size_2015,xiao_modelling_2016,xiao_controlling_2017,deng_continuum_2018} where the segregation force depends on the particle species concentration \cite{van_der_vaart_underlying_2015,jones_asymmetric_2018}. 

The authors acknowledge Adithya Shankar for his contributions to preliminary stages of this work.



%

\end{document}